\documentclass[a4paper,11pt,reqno]{article}

\usepackage{a4wide}
\usepackage{amsmath,amsfonts,amssymb}
\usepackage[english]{babel}
\usepackage{soul}
\usepackage{nicefrac}
\usepackage[mathscr]{euscript}
\usepackage{setspace}
\usepackage{datetime}
\usepackage[sort&compress,merge,numbers]{natbib}

\usepackage{mathrsfs}
\usepackage[T1]{fontenc}
\usepackage{mathpazo}

\usepackage[breaklinks=true]{hyperref}

\numberwithin{equation}{section}

\hypersetup{
			colorlinks=true,
			urlcolor=blue,
			citecolor=magenta,
			linkcolor=blue,
     }

\let\oldsqrt\sqrt
\def\sqrt{\mathpalette\DHLhksqrt}
\def\DHLhksqrt#1#2{%
\setbox0=\hbox{$#1\oldsqrt{#2\,}$}\dimen0=\ht0
\advance\dimen0-0.2\ht0
\setbox2=\hbox{\vrule height\ht0 depth -\dimen0}%
{\box0\lower0.4pt\box2}}

\author{
  \begin{minipage}{.97\linewidth}
    \vspace{1cm}
       \begin{center}
      \begin{small}
               \textbf{Jakob Gath},$^1$ \textbf{Ayan Mukhopadhyay},$^2$
             \textbf{Anastasios C. Petkou},$^3$ 
                     \\
     \textbf{P. Marios Petropoulos}$^1$ and 
      \textbf{Konstantinos Siampos}$^4$
              \end{small}
    \end{center}
    \vspace{0.5cm}
    \hspace{2.4cm}\begin{minipage}{.7\linewidth}
\begin{center}     {\it \begin{footnotesize}
\hbox{\kern-2.4cm\vbox{\vskip0cm
 \begin{itemize}
               \item[$^1$] Centre de Physique Th\'eorique\\ 
        Ecole Polytechnique, CNRS UMR 7644\\
        91128 Palaiseau Cedex, France
  \item[$^3$] Department of Physics\\ 
  Institute of Theoretical Physics\\
  Aristotle University of Thessaloniki\\ 
  54124, Thessaloniki, Greece
\vskip0.31cm
      \end{itemize}}
\kern-4cm\vbox{
\begin{itemize}
  \item[$^2$] Department of Physics\\
  University of Crete\\
  Heraklion 71003, Greece
 \item[$^4$] Albert Einstein Center for Fundamental Physics\\
Institute for Theoretical Physics\\ 
Bern University\\
Sidlerstrasse 5, 3012 Bern, Switzerland
\\
      \end{itemize}\vskip0.05cm
}}
     \end{footnotesize}}
\end{center}
    \end{minipage}
    \vspace{0.5cm}
  \end{minipage}
}

\title{\vspace{3.5cm}
 \boldmath \begin{Large}
    \textbf{PETROV CLASSIFICATION AND HOLOGRAPHIC RECONSTRUCTION OF SPACETIME}
  \end{Large} \unboldmath
}

\date{}

\begin{document}

\begin{titlepage}
\maketitle
\thispagestyle{empty}

 \vspace{-14.5cm}
  \begin{flushright}
  CPHT-RR048.0914\\
  CCQCN-2015-91\\
  CCTP-2015-13
  \end{flushright}
 \vspace{12cm}

\begin{center}
\textsc{Abstract}\\  
\vspace{1cm}	
\begin{minipage}{1.0\linewidth}

Using the asymptotic form of the bulk Weyl tensor, we present an explicit approach that allows us to reconstruct 
exact four-dimensional Einstein spacetimes which are algebraically special with respect to Petrov's classification. 
If the boundary metric supports a traceless, symmetric and conserved complex rank-two tensor, which is related to 
the boundary Cotton and energy--momentum tensors, and if the hydrodynamic congruence is shearless, then the bulk 
metric is exactly resummed and captures modes that stand beyond the hydrodynamic derivative expansion. We illustrate 
the method when the congruence has zero vorticity, leading to the Robinson--Trautman spacetimes of arbitrary 
Petrov class, and quote the case of non-vanishing vorticity, which captures the Pleba\'nski--Demia\'nski Petrov D family. 

\end{minipage}
\end{center}

%\vspace{2cm} 

\end{titlepage}

\onehalfspace

\noindent\rule{\textwidth}{1.2pt}
\vspace{-1cm}
\begingroup
\hypersetup{linkcolor=black}
\tableofcontents
\endgroup
\noindent\rule{\textwidth}{1.2pt}

\section*{Introduction}
\addcontentsline{toc}{section}{Introduction}

Navier--Stokes and Einstein's equations are sets of non-linear equations, which appear to be closely related. This was first noticed in the framework of black holes, where Navier--Stokes describe the dynamics of horizon perturbations \cite{damour1979quelques}. More recently, holography has shed new light in their relationship via the so-called fluid/gravity correspondence in asymptotically anti-de Sitter spacetimes \cite{Hubeny:2011hd}. In this case, fluid dynamics resides on the conformal boundary, and it corresponds to a relativistic conformal fluid described in terms of its traceless and conserved energy--momentum tensor.

The connection between the incompressible horizon fluid and the fluid at the conformal boundary is realized using the holographic renormalization group \cite{EmHuRa, Kuperstein:2013hqa}. Furthermore, an incompressible fluid can also be generically defined in the region between these two extreme points. Its dynamics, \emph{i.e.} the conservation of its energy--momentum tensor, is inherited from the bulk Einstein's momentum constraints, while the Hamiltonian constraint, at leading order of a large-mean-curvature expansion, is interpreted as the equation of state \cite{Bredberg:2010ky, Bredberg:2011jq, Lysov:2011xx}. Such a fluid interpretation is rather formal and may not always be physically accurate since Navier--Stokes equations appear only at first order of a derivative expansion. Moreover, as pointed out in these references, the fluid and gravity degrees of freedom match only under the assumption that the 
Einstein geometry is algebraically special in Petrov's classification. 

The evolution of geometry from the boundary towards the bulk can be formulated as an ADM-type Hamiltonian system which, as usual, requires two pieces of fundamental holographic data.  For pure gravity dynamics, one piece is the boundary metric and the other one is the energy--momentum tensor. If the boundary system is in the hydrodynamic regime, the energy--momentum tensor describes a conformal, non-perfect fluid, but this needs not be true in general for the Hamiltonian evolution scheme to hold. Irrespective of its physical interpretation, the boundary metric together with the energy--momentum tensor  allows us to reconstruct the Einstein bulk spacetime.   

The boundary metric and the boundary energy--momentum tensor are read off in the \emph{Fefferman--Graham expansion} of the bulk metric, as leading and subleading terms, respectively \cite{PMP-FG1, PMP-FG2}. In principle, given the two independent pieces of boundary data the bulk can be reconstructed order by order using the Fefferman--Graham series.  Alternatively, this reconstruction can be achieved with the help of a \emph{derivative expansion}. The latter was originally proposed in \cite{Haack:2008cp, Bhattacharyya:2008jc, 
Bhattacharyya:2008ji}, and is based on the black-brane paradigm. From the bulk perspective, it assumes the existence of a null geodesic congruence defining tubes that extend from the boundary inwards. 
On the boundary, this congruence translates into a timelike congruence, and the aforementioned derivative series expansion is built on increasing derivative order of this field. At the perturbative level, the fluid interpretation is applicable and the boundary timelike congruence is always identified with the boundary fluid velocity field. Beyond the perturbative framework, however, this interpretation is not faithful due to the presence of non-hydrodynamic modes in the boundary energy--momentum tensor.

In general, from a boundary-to-bulk perspective,  
it is unlikely that one could explicitly resum either expansion --  the Fefferman--Graham or the derivative -- 
and the generic bulk solution can be achieved only in a perturbative manner.\footnote{Lorentzian AdS/CFT requires also the careful treatment of initial data at timelike infinities (for a recent mathematical review see \cite{Enciso:2015qva}). This is related to the general discussion of black-hole formation, which is still not fully understood. Since we deal here with exact non-singular bulk solutions, such intricacies are not relevant for us presently.} It makes however sense to pose the following question: given a class of boundary metrics, what are the conditions it should satisfy, and which energy--momentum tensor should it be accompanied with in order for an \emph{exact} dual bulk Einstein space to exist? 

The aim of the present note is to provide a constructive answer to the above question in the case of four-dimensional Einstein spaces. Of course, at this stage, one may wonder why an answer should even exist. Actually, the resummability of the derivative expansion, irrespective of the dimension, was observed in the original papers \cite{Bhattacharyya:2008jc} for the Kerr black holes. This property was latter shown to hold more systematically in four dimensions, even in the presence of a nut charge, which accounts for asymptotically locally anti-de Sitter spacetimes \cite{Papadimitriou:2005ii}. This is achieved by including an infinite, though resummable series of terms built on the boundary Cotton tensor \cite{Caldarelli:2012cm, Mukhopadhyay:2013gja}. There, the requirement was that  the Cotton tensor of the boundary metric be proportional to the energy--momentum tensor, itself being of a perfect-fluid form. This kind of  ansatz unifies all known black-hole solutions with nut charge and rotation, and even allows us to find some new ones (in spirit, this is what happens e.g. when imposing curvature self-duality in Euclidean gravity or in Yang--Mills theories).
It is not expected, however, to exhaust all possibilities, and many Einstein spaces with a rich holographic content are left aside. It is therefore reasonable to attempt finding a generic pattern that guarantees the existence of a bulk dual for appropriately chosen sets of boundary data. 

Here, we will present a general boundary ansatz, which gives us more exact solutions of Einstein's equations in the bulk. Remarkably, we are able to show that our ansatz unifies the above quoted black-hole solutions, described by perfect fluids, with e.g. Robinson--Trautman solutions, whose holographic dual is highly far from equilibrium. Resummation generates therefore non-perturbative effects \emph{i.e.} non-hydrodynamic modes.
The common feature of all these solutions is that they are algebraically special with respect to Petrov's classification: within the proposed method, the Weyl tensor of the four-dimensional bulk is controlled from the boundary data, and turns out to be always at least of type II.

The implications of our work are threefold. Firstly, Einstein's equations are generically non-integrable and the above procedure aims at unravelling integrable sectors in the phase space of solutions, based on appropriate mappings onto integrable dynamical sectors in the dual field theory, such as integrable configurations of Euler's equations for relativistic fluids. Secondly, such a mapping may provide a powerful solution-generating technique, as opposed to standard Geroch-like methods valid in the presence of isometries, which are of limited use in asymptotically anti-de Sitter spaces (see \cite{Leigh:2014dja} for a recent attempt). Thirdly, we can derive an large amount of non-trivial information about holographic strongly coupled field theories: for example in Ref. \cite{Mukhopadhyay:2013gja} it was shown that the existence of exact solutions with perfect-fluid like equilibrium in the perfect-Cotton boundary geometries implied that infinitely many transport coefficients of a special kind should vanish in the dual field theories. 
Enlarging the class of exact solutions with a specific relationship between the boundary data, automatically enables us to obtain highly non-trivial information of multi-point thermal correlation functions of the energy--momentum tensor, even far from the hydrodynamic regime. 

The organisation of the paper is as follows. In the first section we present our ansatz for shaping the boundary data in a manner that guarantees the resummability of the derivative expansion. The relationship between the bulk Weyl tensor and the boundary Cotton and energy--momentum tensors is also clarified. Our approach is constructive and duly motivated, but the formal proof answering the question raised above is left aside and will appear in a separate publication. Instead, we illustrate it in Sec. \ref{exam} by constructing an exact four-dimensional solution of Einstein's equations step by step from an appropriate set of boundary data.  

\section{Bulk reconstruction from boundary data} \label{der-res}

\subsection{The boundary quantities}

Consider a three-dimensional spacetime playing the role of the boundary, equipped with a metric $\text{d}s^2=g_{\mu\nu}\text{d}x^\mu \text{d}x^\nu$ ($\mu, \nu, \ldots = 0,1,2$) and with a symmetric, traceless and covariantly conserved tensor $\text{T}=T_{\mu\nu}\text{d}x^\mu \text{d}x^\nu$. We assume for this tensor the least requirements for being a conformal energy--momentum  tensor \cite{hawking1975large}, and consider systems for which it can be put in the form
\begin{equation}
\label{Tdec}
\text{T}=\text{T}_{(0)}+\Pi  
\end{equation}
with the a perfect-fluid part
\begin{equation}
\label{Tperf}
\text{T}_{(0)}=\frac{\varepsilon}{2}\left(3\text{u}^2+\text{d}s^2\right) \ .
\end{equation}
The timelike congruence $\text{u}=u_\mu(x) \text{d}x^\mu$ is normalized ($u_\mu u^\mu = -1$) and defines the fluid lines. The tensor $\Pi$ captures \emph{all} corrections to the perfect-fluid component, \emph{i.e.} hydrodynamic and non-hydrodynamic modes. The hydrodynamic part is the viscous fluid contribution, which can be expressed as a series expansion with respect to derivatives of $\text{u}$. 
The first derivatives of the velocity field are canonically decomposed in terms of the acceleration $\text{a}$, the expansion $\Theta$, the shear $\sigma$ and the vorticity\footnote{Reminder: $a_\mu=u^\nu\nabla_\nu u_\mu$, $\Theta=\nabla_\mu u^\mu$, $\sigma_{\mu\nu} =  \nabla_{(\mu} u_{\nu )} + a_{(\mu} u_{\nu )} -\frac{1}{2} \Delta_{\mu\nu } \nabla_\rho  u^\rho$, with $\Delta^{\mu\nu} = u^\mu u^\nu + g^{\mu\nu}$ the projector onto the space orthogonal to $\text{u}$.}
\begin{equation}\label{}
\omega=\frac{1}{2}\omega_{\mu\nu }\, \mathrm{d}\mathrm{x}^\mu\wedge\mathrm{d}\mathrm{x}^\nu  =\frac{1}{2}\left(\mathrm{d}\mathrm{u} +
\mathrm{u} \wedge\mathrm{a} \right).
\end{equation}

In the Landau frame, the hydrodynamic component of $\Pi$ is transverse to $\text{u}$. The full $\Pi$ \emph{is not} transverse but 
\begin{equation}
\label{Piprop}
\Pi_{\mu\nu}u^\mu u^\nu=0 \quad \Rightarrow \quad T_{\mu\nu}u^\mu u^\nu=\varepsilon(x).
\end{equation}
The latter is the local energy density, related to the pressure via the conformal equation of state $\varepsilon=2p$. However, it should be stressed that the presence of a non-hydrodynamic component tempers the fluid interpretation. In particular, it is not an easy  task to extract the congruence $\text{u}$, because its 
meaning as a vector tangent to fluid lines becomes questionable. 

Another important structure in three spacetime dimensions, where the Weyl tensor vanishes, is the Cotton tensor\footnote{The Cotton and Levi--Civita are pseudo--tensors, \emph{i.e.} they change sign under a parity transformation. It is therefore important  to state the convention in use.}
\begin{equation}
\label{cotdef}
C^{\mu\nu}=\eta^{\mu\rho\sigma}
\nabla_\rho \left(R^{\nu}_{\hphantom{\nu}\sigma}-\frac{R}{4}\delta^{\nu}_{\hphantom{\nu}\sigma} \right) 
\end{equation}
with $\eta^{\mu\nu\sigma}=\nicefrac{\epsilon^{\mu\nu\sigma}}{\sqrt{-g}}$. This tensor vanishes if and only if the spacetime is conformally flat. It shares the key properties of the energy--momentum tensor, \emph{i.e.} it is symmetric, traceless and covariantly conserved. For later reference we introduce a contraction analogous to the energy density \eqref{Piprop},
\begin{equation}
\label{cofx}
C_{\mu \nu}u^\mu u^\nu = c(x)\ .
\end{equation}

\subsection{Bulk Petrov classification and the resummability conditions} \label{Pet}

The four-dimensional Weyl tensor can be classified into distinct types, \emph{i.e.} according to the algebraic Petrov types. For an Einstein space (with a given sign of the Ricci curvature) this provides a complete classification of the curvature tensor.

In order to establish a connection with the three-dimensional boundary data it is useful to recall how the algebraic Petrov classification is obtained from the eigenvalue equation for the Weyl tensor. 
In particular, the Weyl tensor and its dual can be used to form a pair of complex-conjugate tensors.
Each of these tensors has two pairs of bivector indices, which can be used to form a complex two-index tensor. Its components are naturally packaged inside a complex symmetric $3 \times 3$ matrix $\text{Q}$ with zero trace (see e.g. \cite{Stephani:624239} for this construction).
This matrix encompasses the ten independent real components of the Weyl tensor and the associated eigenvalue equation determines the Petrov type.

Performing the Fefferman--Graham expansion of the complex Weyl tensor $\text{Q}^{\pm}$ for a general Einstein space, one can show that the leading-order ($\nicefrac{1}{r^3}$) coefficient, say $\text{S}^{\pm}$, exhibits a specific combination of the components of the boundary Cotton and energy--momentum tensors.\footnote{We will provide the details in the already announced upcoming publication (see also e.g. \cite{Mansi:2008br, Mansi:2008bs}).} 
The algebraic Segre type of this combination determines precisely the Petrov type of the four-dimensional bulk metric and establishes a one-to-one map between the bulk Petrov type and the boundary data.

Assume now that we wish to reconstruct the Einstein bulk spacetime from a set of boundary data.
Given a three-dimensional boundary metric, one can \emph{impose} a desired \emph{canonical form} for the asymptotic Weyl tensor $\text{S}^{\pm}$, as e.g. a perfect-fluid form (type D) or matter--radiation form (type III or N) or a combination of both (type II) (see e.g. \cite{Chow:2009km} for these structures).
Doing so, \emph{not only do we design from the boundary the special algebraic structure of the bulk spacetime, but we also provide a set of conditions that turn out to guarantee the resummability of the perturbative expansion into an exact Einstein space.} This is our central result as it answers the question asked earlier in the introduction. The rest of the paper will be devoted to making this statement as clear as possible and illustrating it with robust examples. 

It turns out that it is somehow easier to work with a different pair of complex-conjugate tensors
\begin{equation}
	\label{eqn:Tref}
	T_{\mu\nu}^\pm = T_{\mu\nu} \pm \frac{i}{8\pi G k^2 }C_{\mu\nu} \ ,
\end{equation}
{where $k$ is a constant
and $\text{T}^\pm$ is related to  $\text{S}^{\pm}$ by a similarity transformation: 
$\text{T}^\pm =- \text{P} \, \text{S}^{\pm}\text{P}^{-1}$ with $\text{P}={\rm diag}(\pm i,-1,1)$.} 
Choosing a specific canonical form for these tensors, and assuming a boundary metric $\text{d}s^2$, we are led to two conditions. The first, provides a set of equations that the boundary metric must satisfy:
\begin{equation}
\label{C-con}
\boxed{
\text{C}=8\pi G k^2\,  \text{Im} \text{T}^+.}
    \end{equation}
The second delivers the boundary energy--momentum tensor it should be accompanied with for an exact bulk ascendent spacetime to exist:
\begin{equation}
\label{T-con}
\boxed{\text{T}= \text{Re} \text{T}^+.}
    \end{equation}
The tensors given in Eq.~\eqref{eqn:Tref} are by construction symmetric, traceless and conserved:
\begin{equation}
\label{Tref-cons}
\boxed{\nabla\cdot \text{T}^\pm=0.}
    \end{equation}
 We will refer to them as the \emph{reference energy--momentum tensors} as they play the role of a pair of fictitious conserved boundary sources, always accompanying the boundary geometry. 
It turns out that the particular combination \eqref{eqn:Tref} of the energy--momentum and Cotton tensors is exactly the combination one finds if the  Weyl tensor is decomposed into self-dual and anti-self-dual components, which given the Lorentzian signature are complex-conjugate. These are nicely captured in the Cahen--Debever--Defrise\footnote{The decomposition is more commonly known as Atiyah--Hitchin--Singer.} decomposition.

Finally, we note that some care must be taken when working with $\text{T}^\pm$ instead of $\text{S}^\pm$. 
Indeed, the eigenvalues are equal, but not necessarily their eigenvectors. In particular, this means that 
one cannot determine the Petrov type unambiguously if considering the eigenvalue equation for 
$\text{T}^{\pm}$.\footnote{The ambiguities are between type II (III) and type D (N), since these types have the 
same degeneracy of eigenvalues. This will be noticed in the example analyzed in Sec. \ref{RRT}.}

\subsection{The derivative expansion and its ressumation ansatz}

We have listed in the previous section all boundary ingredients needed for reaching holographically exact bulk Einstein spacetimes. We would like here to discuss their actual reconstruction. We will use for that the derivative expansion, organized around the derivatives of the boundary fluid velocity field $\text{u}$. This expansion assumes small derivatives, small curvature, and small higher-derivative curvature tensors for the boundary metric. This limitation is irrelevant for us since we are ultimately interested in resumming the series. A related and potentially problematic issue, is the definition of $\text{u}$, which is not automatic when the boundary energy--momentum tensor $\text{T}$ is not of the fluid type. In that case $\text{u}$ should be considered as an extra ingredient of the ansatz, \emph{a posteriori} justified by the success of the resummation.

The guideline for the reconstruction of spacetime  based on the derivative expansion is \emph{Weyl covariance} \cite{Haack:2008cp, Bhattacharyya:2008jc}: the bulk geometry should be insensitive to a conformal rescaling of the boundary metric $\text{d}s^2\to  \nicefrac{\text{d}s^2}{{\cal B}^2}$. The latter is accompanied with $\text{C}\to {\cal B}\, \text{C}$, and at the same time $\text{T}\to {\cal B}\, \text{T}$, $\text{u}\to \nicefrac{\text{u}}{{\cal B}}$ (velocity one-form) and $\omega\to \nicefrac{\omega}{{\cal B}}$ (vorticity two-form). Covariantization with respect to rescalings requires to introduce a Weyl connection one-form:
\begin{equation}
\label{Wcon}
\text{A}:=\text{a} -\frac{\Theta}{2} \text{u} \ ,
\end{equation}
which transforms as $\text{A}\to\text{A}-\text{d}\ln {\cal B}$. Ordinary covariant derivatives $\nabla$ are thus traded for Weyl-covariant ones $\mathscr{D}=\nabla+w\,\text{A}$, $w$ being the conformal weight of the tensor under consideration. In three spacetime dimensions, Weyl-covariant quantities are e.g. 
\begin{eqnarray}
\mathscr{D}_\nu\omega^{\nu}_{\hphantom{\nu}\mu}&=&\nabla_\nu\omega^{\nu}_{\hphantom{\nu}\mu},\\
\mathscr{R}&=&R +4\nabla_\mu A^\mu- 2 A_\mu A^\mu \ ,
\end{eqnarray}
while
\begin{equation}
\label{sigma}
\Sigma=
\Sigma_{\mu\nu} 
\text{d}x^\mu\text{d}x^\nu=-2\text{u}\mathscr{D}_\nu \omega^\nu_{\hphantom{\nu}\mu}\text{d}x^\mu- \omega_\mu^{\hphantom{\mu}\lambda} \omega^{\vphantom{\lambda}}_{\lambda\nu}\text{d}x^\mu\text{d}x^\nu
-\text{u}^2\frac{\mathscr{R}}{2} \ ,
\end{equation}
is Weyl-invariant. Notice also that for any symmetric and traceless tensor $S_{\mu\nu}\text{d}x^\mu\text{d}x^\nu$ of conformal weight $1$ (like the energy--momentum tensor and the Cotton tensor) has
\begin{equation}
\mathscr{D}_\nu S^{\nu}_{\hphantom{\nu}\mu}=\nabla_\nu S^{\nu}_{\hphantom{\nu}\mu} \ .
\end{equation}

In the present analysis, we will be interested in situations where the boundary congruence $\text{u}$ is \emph{shear-free}. 
Despite this limitation, wide classes of dual holographic bulk geometries remain accessible. 
Vanishing shear simplifies considerably the reconstruction of the asymptotically AdS bulk geometry because it reduces the available Weyl-invariant terms. As a consequence, at each order of $\mathscr{D}\text{u}$, the terms compatible with  Weyl covariance of the bulk metric $\text{d}s^2_{\text{bulk}}$ are nicely organized. Even though we cannot write them all at arbitrary order, the structure of the first orders suggests that resummation, whenever possible, should lead to the following \cite{Haack:2008cp, Bhattacharyya:2008jc, Bhattacharyya:2008ji, Caldarelli:2012cm, Mukhopadhyay:2013gja, Petropoulos:2014yaa}:
\begin{equation}
\text{d}s^2_{\text{res.}} =
-2\text{u}(\text{d}r+r \text{A})+r^2k^2\text{d}s^2+\frac{\Sigma}{k^2}
+ \frac{\text{u}^2}{\rho^2} \left(\frac{3 T_{\lambda \mu}u^\lambda u^\mu}{k \kappa }r+\frac{C_{\lambda \mu}u^\lambda \eta^{\mu\nu\sigma}\omega_{\nu\sigma}}{2k^6}\right).
\label{papaefgenres}
\end{equation}
Here $r$ the radial coordinate whose dependence is explicit, 
$x^\mu$ are the three boundary coordinates extended to the bulk, on which depend implicitly the various functions,
$\eta^{\mu\nu\sigma}=\nicefrac{\epsilon^{\mu\nu\sigma}}{\sqrt{-g}}$, $\kappa=\nicefrac{3k}{8\pi G}$, $k$ a constant,
and $\Sigma$ is displayed in \eqref{sigma}. Finally,\footnote{The three-dimensional Hodge dual of the vorticity is always aligned with the velocity field and this is how $q(x)$ is originally defined:
$\eta^{\mu\nu\sigma}\omega_{\nu\sigma}=q u^\mu$.} 
\begin{equation}\label{rho2}
 \rho^2=r^2 +\frac{1}{2k^4} \omega_{\alpha\beta} \omega^{\alpha\beta} := r^2 +\frac{q^2}{4k^4}
\end{equation}
performs the resummation as the derivative expansion is manifestly organized in powers of $q^2=2 \omega_{\alpha\beta} \omega^{\alpha\beta}$. This structure is inferred by the first orders, which are the ones that have been explicitly determined in Refs. \cite{Bhattacharyya:2008jc, Caldarelli:2012cm}.
In expression \eqref{papaefgenres}, we recognize the energy density $\varepsilon(x)$ introduced in \eqref{Piprop}, and 
 $c(x)$ as in \eqref{cofx}. 
The presence of the boundary Cotton tensor stresses that the bulk is generically  asymptotically \emph{locally} anti-de Sitter. It is readily checked that boundary Weyl transformations correspond to bulk diffeomorphisms, which can be reabsorbed into a redefinition of the radial coordinate: $r\to {\cal B}\, r$.

The four-dimensional metric $\text{d}s^2_{\text{res.}}$ displayed in \eqref{papaefgenres} is not expected to be Einstein for arbitrary boundary data $\text{T}$ and $\text{d}s^2$. Our claim is that when these data satisfy Eqs. \eqref{C-con} and \eqref{T-con},  $\text{d}s^2_{\text{res.}}$ is Einstein with $\Lambda=-3 k^2$. 
 Following the discussion of Sec. \ref{Pet}, \emph{this spacetime is algebraically special}, its Petrov type being determined directly by the \emph{a priori} chosen reference tensor $\text{T}^\pm$ subject to \eqref{Tref-cons}, necessary for the conditions  \eqref{C-con} and \eqref{T-con} to be used. Hence, \emph{Eqs.  \eqref{C-con} and \eqref{Tref-cons} should be considered as a boundary translation of Einstein's equations}, in some integrable sector of algebraically special geometries. This is the central message of the present work. Scanning over canonical forms for $\text{T}^+$ amounts to exploring various Petrov classes.
  Hence, the metric \eqref{papaefgenres} admits degenerate principal null directions and it is thus of type II,  III,  N, D or O. 

Several remarks are in order here. Being algebraically special, the spacetimes at hand must admit a null, geodesic and shear-free congruence, as stated in the Goldberg--Sachs theorem. The congruence $\text{u}$ in the bulk is null and geodesic, and becomes timelike and shear-free (but not longer necessarily geodesic) on the boundary, where it identifies with the fluid velocity field. It turns out to be indeed shear-free everywhere in the bulk, provided the conditions \eqref{C-con} and  \eqref{T-con} are fulfilled. The absence of shear for the boundary fluid congruence seem therefore to be intimately related to the resummability of the derivative expansion into an algebraically special Einstein space. This is in agreement with the fact that the large number of Weyl-covariant tensors available when the shear is non-vanishing, makes it unlike that the resummation occurs.

As already stressed previously, the energy--momentum tensor $\text{T}$, obtained in the procedure described in Sec. \ref{Pet}, is not necessarily of the fluid type (we shall soon meet examples in Sec. \ref{exam}). Hence, it is not straightforward to extract the velocity congruence $\text{u}$, required in the resummed expression \eqref{papaefgenres} -- and further check or impose the absence of shear. The determination of a shearless 
$\text{u}$ should therefore be considered as part of the ansatz. Assume for concreteness a
 boundary metric of the form
\begin{equation}
\label{PDbdymet}
\text{d}s^2=-\Omega^2(\text{d}t-\text{b})^2+\frac{2}{k^2P^2}\text{d}\zeta\text{d}\bar\zeta,
\end{equation}
where $P$ and $\Omega$ are arbitrary real functions of $(t,\zeta, \bar \zeta)$, and 
\begin{equation}
\label{frame}
\text{b}=B(t,\zeta, \bar \zeta)\, \text{d}\zeta+\bar B(t,\zeta, \bar \zeta)\, \text{d}\bar\zeta.
\end{equation}
This is actually the most general three-dimensional metric,\footnote{We could even set $\Omega=1$, without spoiling the generality.} as we make no assumption regarding isometries, with a specific choice of local frames.
Part of our resummation ansatz is to assume that the boundary frame has been adapted to the fluid shear-free congruence, so that 
\begin{equation}
\label{ut}
\text{u}= -\Omega(\text{d}t-\text{b}).
\end{equation}
We can thus express the resummed four-dimensional bulk metric \eqref{papaefgenres} in terms of a null tetrad as 
\begin{equation}
\label{papaefgentetr}
\boxed{
\text{d}s^2_{\text{res.}} =-2\mathbf{k}\mathbf{l}+2\mathbf{m}\bar{\mathbf{m}},}
\end{equation}
where
\begin{equation}
\label{km}
\mathbf{k}=-\text{u},\quad
\mathbf{m}=\frac{\rho}{P}\text{d}\zeta
\end{equation}
and\footnote{The Hodge duality is here meant with respect to the three-dimensional boundary:
$\ast(\text{u}\wedge \text{d} q)= \eta_{\mu}^{\hphantom{\mu}\nu\sigma}u_\nu\partial_\sigma q\, \text{d}x^\mu$.}  
\begin{equation}
\label{l}
\mathbf{l}=-\text{d}r-r \text{a} -H \text{u}
+
\frac{1}{2k^2} \ast(\text{u}\wedge (\text{d} q+q\text{a}))
\end{equation}
with
\begin{equation}
\label{Hgen}
2 H= r^2k^2 -r\, \Theta
+\frac{q^2}{k^2}
+\frac{\mathscr{R}}{2k^2}
-\frac{3 }{\rho^2 k} \left(\frac{r\varepsilon}{\kappa}+\frac{qc}{ 6k^5}
\right).
\end{equation}
In the latter expression we have introduced
 $\varepsilon(x)$ and $c(x)$ defined in \eqref{Piprop} and \eqref{cofx} ($x$ refers to the coordinates $t,\zeta,\bar\zeta$ common for bulk and boundary).

\section{Concrete examples} \label{exam}

\subsection{The boundary metric and the reference energy--momentum tensors}

The resummation method presented here generalizes previous successful attempts to reconstruct exact Einstein spaces from boundary data  \cite{Caldarelli:2012cm, Mukhopadhyay:2013gja}. In these works, the boundary metric  was of the type \eqref{PDbdymet} with two commuting Killing vectors, and the energy--momentum tensor was perfect-fluid and proportional to the Cotton tensor. This is a particular case of our present ansatz, with hydrodynamic boundary state
{(see Sec. \ref{vort})}. We will now move to a different situation and consider a specific family of boundary geometries, namely those with
\begin{equation}
\label{RTbdymet}
\text{d}s^2=-\text{d}t^2+\frac{2}{k^2P^2}\text{d}\zeta\text{d}\bar\zeta.
\end{equation}
This is not the most general three-dimensional metric because it follows from \eqref{PDbdymet} with $\Omega=1$ and $\text{b}=0$. As we will see soon, it turns out to enable the holographic reconstruction of Robinson--Trautman Einstein metrics of all Petrov types. Here 
the Cotton tensor, computed using \eqref{cotdef}, reads:
\begin{equation}
\label{cot}
\text{C}=-i \left(\begin{matrix}
\text{d}t& \text{d}\zeta & \text{d}\bar\zeta
\end{matrix}\right)
\left(\begin{matrix}
 0&-\frac{k^2}{2}\partial_\zeta K&\frac{k^2}{2}\partial_{\bar\zeta} K\\
-\frac{k^2}{2}\partial_\zeta K &-\partial_t\left(\frac{\partial^2_\zeta P}{P}
\right) &0 \\
\frac{k^2}{2}\partial_{\bar\zeta} K&0&\partial_t\left(\frac{\partial^2_{\bar\zeta} P}{P}\right) 
 \end{matrix}\right) \left(\begin{matrix}
\text{d}t \\ \text{d}\zeta \\ \text{d}\bar\zeta
\end{matrix}\right),
\end{equation}
where 
\begin{equation}
K=2P^2\partial_\zeta\partial_{\bar\zeta}\log P
\end{equation}
is the Gaussian curvature of the surfaces at constant $t$ divided by $k^2$. This tensor is real.

We must now introduce a canonical reference energy--momentum tensor $\text{T}^\pm$ and apply the strategy 
displayed above: (\romannumeral1) impose conservation \eqref{Tref-cons}; (\romannumeral2) constrain the
boundary metric using  \eqref{C-con} and determine the actual energy--momentum tensor with \eqref{T-con}; 
(\romannumeral3) reconstruct the bulk Einstein space using \eqref{papaefgentetr}. Equations  \eqref{Tref-cons}
and  \eqref{C-con} are expected to guarantee that Einstein's equations are fulfilled and provide information on the reached Petrov class. The latter must be related to the choice of reference tensor $\text{T}^\pm$. 
There are two basic and distinct cases will be considered here.

\paragraph{Perfect-fluid form.} For perfect-fluid reference tensors, it is necessary to introduce two complex-conjugate reference velocity congruences $\text{u}^{\pm}$. It is not useful to analyze the most general velocity congruences, but the most typical ones, keeping in mind that a redundancy is expected to exist, making different-looking boundary data correspond to identical Einstein spaces. Consider the normalized reference congruence\footnote{These are the most general ones: adding an extra leg along the missing direction, and adjusting the overall scale for keeping the norm to $-1$ amounts to a combination of a diffeomorphism and a Weyl transformation.}
 \begin{equation}
 \label{upm}
 \text{u}^+= \text{u}+ \frac{\alpha^+}{P^2}\text{d}\zeta
  \end{equation}
with (see \eqref{ut})
 \begin{equation}
 \label{uphys}
 \text{u}=-\text{d}t
  \end{equation}
the physical congruence,
$\alpha^+= \alpha^+(t,\zeta,\bar \zeta)$, and its complex-conjugate $\text{u}^-= \text{u}+ \frac{\alpha^-}{P^2}\text{d}\bar \zeta$ with $\alpha^-=\alpha^{+\ast}$. The physical congruence $\text{u}$ is shear-free, has no vorticity, no acceleration but is expanding at a rate
\begin{equation}
\Theta=-2 \partial_t \log P.
\end{equation}

A perfect-fluid energy-momentum tensor based on these reference congruences reads:
\begin{equation}
\label{RT-perflu}
\text{T}^\pm_{\text{pf}}= \frac{M_\pm(t,\zeta,\bar \zeta) k^2}{8\pi G}\left(3\left(\text{u}^\pm\right)^2 +\text{d}s^2\right)
\end{equation}
with $M_-=M_+^\ast$.
The choice of the functions $\alpha^\pm(t,\zeta,\bar \zeta)$ and $M_\pm (t,\zeta,\bar \zeta)$ should be restricted so that $\text{T}^\pm$ is conserved \emph{i.e.} \eqref{Tref-cons}
is fulfilled.

 At this stage we may pose and ask a generic question: given a velocity congruence with expansion and acceleration, can one find a pressure field such that the corresponding traceless 
perfect-fluid energy--momentum tensor is conserved? The analysis of that question is performed in App. \ref{appendix.perfect} and the answer is the following: a pressure locally exists if and only if the Weyl connection constructed out of the velocity, the expansion and the acceleration is flat (zero exterior derivative). If furthermore the Weyl connection is vanishing, the pressure is a constant.
Here
\begin{equation}
\label{Am}
\text{A}^-=\text{a}^- -\frac{\Theta_-}{2} \text{u}^-
=\frac{\Theta_-}{2} \text{d}t+\left(\frac{\Theta_-\alpha^-}{2P^2}+\frac{\partial_t\alpha^-}{P^2}\right)\text{d}\bar \zeta
\end{equation}
with
\begin{equation}
\label{thetm}
\Theta_-
=k^2 \partial_\zeta\alpha^-- 2\left(k^2\alpha^- \frac{\partial_\zeta P}{P}+ \frac{\partial_t P}{P}\right),
\end{equation}
and similarly for the ``+'' by complex conjugation. We must impose  $\text{d}A^{\pm}=0$ and determine the reference pressures $p_\pm(t,\zeta,\bar \zeta)$ such that $A^{\pm}=\text{d}\ln p_\pm^{\nicefrac{-1}{3}}$.
The closure of $A^{\pm}$ can be worked out systematically, with a simple generic solution. Using \eqref{Am} 
we find that the functions $\alpha^\pm$ must be factorized:
\begin{equation}
\label{fact}
\alpha^\pm(t,\zeta,\bar \zeta) = g_\pm(t)\,  \tilde \alpha^\pm(\zeta,\bar \zeta)
\end{equation}
with 
\begin{equation}
\Theta_\pm
= - 2\partial_t \ln g_\pm.
\end{equation}
Using \eqref{thetm}, the latter can be written explicitly as 
\begin{equation}
\boxed{ \partial_\zeta\alpha^-+\frac{2}{k^2} \partial_t \ln g_\pm= 2\left(\alpha^- \frac{\partial_\zeta P}{P}+ \frac{\partial_t P}{k^2P}\right)\quad \text{and\quad c.c. }.}\label{h}
\end{equation}
In conclusion, a congruence solving Euler equations is characterized by two  functions and their complex conjugates:  $\tilde \alpha^\pm(\zeta,\bar\zeta)$ and  
\begin{equation}
p_\pm (t) =  \frac{M_\pm(t) k^2}{8\pi G}=  g^3_\pm(t).
\end{equation}
The product  \eqref{fact} must satisfy Eq. \eqref{h}.

\paragraph{Radiation-matter form.} 

Consider finally
\begin{equation}
\label{rmt}
8\pi G \text{T}^{+}_{\text{rm}}=2
\text{d}\zeta\left(\beta \text{d}t+\frac{\gamma}{k^2} \text{d}\zeta\right).
\end{equation}
In this expression $\beta$ and $\gamma$ are \emph{a priori} functions of $t,\zeta$ and $\bar \zeta$. The tensor is the symmetrized  direct product of a light-like by a time-like vector. Its conservation enforces the dependence $\beta=\beta(t,\zeta)$ and the condition
\begin{equation}
\label{rmth}
\boxed{\partial_t \beta
-\beta \partial_t \ln P^2
-2P^2
\partial_{\bar\zeta}\gamma=0.}
\end{equation}
Notice that for vanishing $\beta$, we obtain a \emph{pure-radiation} tensor \emph{i.e.} the square of of a null vector. In this case, the conservation equation \eqref{rmth} enforces the reduced dependence $\gamma=\gamma(t,\zeta)$.

\subsection{Resummation: the Robinson--Trautman Einstein spaces}\label{RRT}

The preceding analysis has not put any restriction on the boundary metric \eqref{RTbdymet}. It simply selected 
the appropriate ingredients for a reference tensor to be conserved. We will consider a general conserved reference tensor of the form
\begin{equation}
\label{gent}
\text{T}^{+}=
\text{T}^{+}_{\text{pf}}+\text{T}^{+}_{\text{rm}},
\end{equation}
the two components being given in Eqs. \eqref{RT-perflu} and \eqref{rmt}. For this combination, 
\begin{eqnarray}
\label{T-com}
&&\frac{8\pi G}{i}   \text{Im}\text{T}^+=\nonumber \\&&  \left(\begin{matrix}
\text{d}t& \text{d}\zeta & \text{d}\bar\zeta
\end{matrix}\right)
\left(\begin{matrix}
\displaystyle{ k^2\left(M_--M_+\right)}
 &\frac{3M_+\alpha^+k^2}{2P^2} -\frac{\beta}{2}
 &\frac{-3M_-\alpha^-k^2}{2P^2}+\frac{\bar \beta}{2}
 \\
\frac{3M_+\alpha^+k^2}{2P^2} -\frac{\beta}{2}
&\frac{-3M_+(\alpha^+)^2k^2}{2P^4}-\frac{\gamma}{k^2}&\frac{M_--M_+}{2P^2}
\\
\frac{-3M_-\alpha^-k^2}{2P^2}+\frac{\bar \beta}{2}&\frac{M_--M_+}{2P^2}&
\frac{3M_-(\alpha^-)^2k^2}{2P^4}+\frac{\bar\gamma}{k^2}
\end{matrix}\right) \left(\begin{matrix}
\text{d}t \\ \text{d}\zeta \\ \text{d}\bar\zeta
\end{matrix}\right)\quad
    \end{eqnarray}
and we can now require \eqref{C-con}. The first observation is that this identification of the Cotton tensor requires 
\begin{equation}
\label{bondi}
M_+(t)=M_-(t),
\end{equation}
which we will name $M(t)$, a real function. Furthermore,  it appears 
a pair of independent conditions plus their complex-conjugates. The first reads:
\begin{equation}
\boxed{ \partial_t\left(\frac{\partial^2_{\zeta} P}{P}\right) +\frac{3}{2}Mk^4\frac{(\alpha^+)^2}{P^4}+\gamma=0\quad \text{and\quad c.c. },}\label{h3}
\end{equation}
while the second is 
\begin{equation}
\boxed{\partial_{\zeta} K+\beta=3Mk^2\frac{\alpha^+}{P^2}\quad \text{and\quad c.c. }.}\label{h2}
\end{equation}

Equations \eqref{h3} and \eqref{h2} are \emph{algebraic constraints} on the functions which determine the boundary energy--momentum tensor: $\alpha^\pm(t,\zeta,\bar \zeta) =\left( \nicefrac{M(t) k^2}{8\pi G}\right)^{\nicefrac{1}{3}}  \tilde \alpha^\pm(\zeta,\bar \zeta)$, $\beta(t,\zeta)$ and  $\gamma(t,\zeta, \bar \zeta)$  (as well as the complex conjugate functions  of $\beta,\gamma$). These algebraic relationships do not constrain the three-dimensional geometry, unless we require by hand some extra conditions on the functions 
$\alpha, \beta,\gamma$. We will come back later to this possibility.

Besides Eqs. \eqref{h3} and \eqref{h2}, we must also impose the differential conditions \eqref{h} and \eqref{rmth}, which guarantee the conservation of the reference energy--momentum tensor \eqref{gent} and consequently of the genuine boundary energy--momentum \eqref{T-con}. These two conditions are actually redundant, once \eqref{h3} and \eqref{h2} are taken into account. This redundancy is due to the fact that the Cotton tensor \eqref{cot} is identically conserved. The resulting unique independent conservation condition obtained by combining e.g.  \eqref{h} with \eqref{h2} reads:
\begin{equation}
\boxed{\Delta K -12M \partial_t \log P+4  \partial_t M=0,}
\label{E}
\end{equation}
where $\Delta = 2P^2 \partial_{\bar\zeta} \partial_\zeta$. 
This is a differential equation for the boundary metric $\text{d}s^2$ given in  \eqref{RTbdymet}, and for $M(t)$.
It should be interpreted as an \emph{integrability condition} for the resummed series expansion \eqref{papaefgenres} to be exactly Einstein. 

We can now proceed and determine the bulk metric $\text{d}s^2_{\text{res.}}$, using  \eqref{papaefgenres}. For that we need the energy--momentum tensor, given in terms of the reference tensor \eqref{gent} by  \eqref{T-con}.  Inserting \eqref{bondi} as well as the algebraic conditions \eqref{h3} and \eqref{h2} in the latter, we obtain the boundary energy--momentum tensor exclusively in terms of the metric data $P,K$ and the function $M(t)$:
\begin{equation}
\label{RT-full}
\text{T}=\frac{1}{16\pi G} \left(\begin{matrix}
\text{d}t& \text{d}\zeta & \text{d}\bar\zeta
\end{matrix}\right)
\left(\begin{matrix}
 4Mk^2&-\partial_\zeta K&-\partial_{\bar\zeta} K\\
-\partial_\zeta K &-\frac{2}{k^2}\partial_t\left(\frac{\partial^2_{\zeta} P}{P}\right)&\frac{2M}{P^2}
\\
-\partial_{\bar\zeta} K&\frac{2M}{P^2}&-\frac{2}{k^2}\partial_t\left(\frac{\partial^2_{\bar\zeta} P}{P}\right) \end{matrix}\right) \left(\begin{matrix}
\text{d}t \\ \text{d}\zeta \\ \text{d}\bar\zeta
\end{matrix}\right).
\end{equation}
This tensor can be put in the form \eqref{Tdec}, \eqref{Tperf} with $\text{u}$ in \eqref{uphys} and $\text{d}s^2$ in \eqref{RTbdymet}. The energy density is determined using \eqref{Piprop} and \eqref{uphys}:
\begin{equation}
\label{endenRT}
\varepsilon(t) = \frac{M(t) k^2}{4\pi G},
\end{equation}
and is
time-dependent. The non-perfect component  reads:
\begin{equation}
\Pi=\frac{1}{16\pi G} \left(\begin{matrix}
\text{d}t& \text{d}\zeta & \text{d}\bar\zeta
\end{matrix}\right)
\left(\begin{matrix}
 0&-\partial_\zeta K&-\partial_{\bar\zeta} K\\
-\partial_\zeta K &-\frac{2}{k^2}\partial_t\left(\frac{\partial^2_{\zeta} P}{P}\right) &0
\\
-\partial_{\bar\zeta} K&0&-\frac{2}{k^2}\partial_t\left(\frac{\partial^2_{\bar\zeta} P}{P}\right) 
 \end{matrix}\right) \left(\begin{matrix}
\text{d}t \\ \text{d}\zeta \\ \text{d}\bar\zeta
\end{matrix}\right),
\end{equation}
and contains both hydrodynamic and non-hydrodynamic components. 

Putting everything together (here $\omega$ and $q$ vanish\footnote{Note also that $\mathscr{R}=2k^2 K$ and $\Sigma=-k^2K\text{d}t^2$.}) we obtain \eqref{papaefgentetr} with
\begin{equation}
\label{RT}
\mathbf{k}=-\text{u}, \quad
\mathbf{l}=-\text{d}r -H \text{u}, \quad \mathbf{m}=\frac{r}{P}\, \text{d}\zeta
\end{equation}
and\begin{equation}
\label{H}
2H = k^2 r^2 + 2r \partial_t \log P+  K -\frac{2M}{r} .
\end{equation}
Our claim, according to the analysis in Sec. \ref{der-res}, is that under the condition \eqref{E}, Eq.  \eqref{papaefgenres} with \eqref{RT} and \eqref{H} provides an exact Einstein space. This turns out to be precisely the case, and the reader might have recognized in Eqs.  \eqref{papaefgenres}, \eqref{RT} and \eqref{H} the Robinson--Trautman ansatz, which is Einstein precisely under the condition \eqref{E}.\footnote{Holography in Robinson--Trautman spacetimes has recently attracted some attention \cite{deFreitas:2014lia,Bakas:2014kfa}. In particular, the holographic energy--momentum tensor found from the bulk geometry in \cite{deFreitas:2014lia} agrees with ours (Eq. \eqref{RT-full}), obtained from purely boundary considerations.}  This condition, found in our method from purely boundary considerations
is nothing but bulk Einstein's equation, and $M(t)$ the Bondi mass. This result is remarkable because it shows, as anticipated, that Eq.~\eqref{C-con} together with the conservation of the reference tensor \eqref{gent}  are indeed, from the boundary, equivalent to Einstein's equations in some integrable sector. 

In order to close this analysis, we would like to come back to the two algebraic equations. \eqref{h3} and \eqref{h2} that the functions entering the boundary energy--momentum tensor should satisfy. In order to clarify their role, it is appropriate to remind that the bulk Robinson--Trautman metric is algebraically special, \emph{i.e.} generically Petrov type II. Choosing the bulk null tetrad as in \eqref{RT}, the non-vanishing components of the Weyl tensor are 
\begin{eqnarray}
\label{Psi2}\Psi_2&=&-\frac{M}{r^3},\\
\label{Psi3}\Psi_3&=&-\frac{P}{2 r^2}  \partial_\zeta K,\\
\label{Psi4}\Psi_4&=&\frac{1}{2 r^2}  \partial_\zeta \left(P^2 \partial_\zeta K\right)+\frac{P^2 }{r}
 \partial_t\left(\frac{\partial^2_\zeta P}{P}
 \right). 
\end{eqnarray}
The direction $\mathbf{k}$, which on the boundary becomes the time-like congruence $\text{u}$, is generically a \emph{doubly degenerate principal null direction} because the conditions \eqref{h3} and  \eqref{h2}  leave enough freedom on the \emph{a priori} arbitrary functions 
$M(t)$, $\tilde \alpha^\pm(\zeta,\bar \zeta)$, $\beta(t,\zeta)$ and $\gamma(t,\zeta, \bar \zeta)$ to avoid any constraint on the functions $P(t,\zeta, \bar \zeta)$ or $K=\Delta \ln P$.

We may however tune the various functions defining the reference energy--momentum tensor \eqref{gent}, in order to increase the degeneracy of the bulk principal null direction, and explore in a \emph{boundary-controlled} manner other Petrov bulk geometries.
\begin{itemize}
\item Set $M(t)=0$. This amounts to keeping\footnote{Note that for vanishing $M(t)$, the functions $\tilde \alpha^\pm(\zeta,\bar\zeta)$ become irrelevant.} a purely radiation-matter reference energy--momentum tensor. Now \eqref{h3} reads:
\begin{equation}
 \partial_t\left(\frac{\partial^2_{\zeta} P}{P}\right) +\gamma=0\quad \text{and\quad c.c. },\label{h3III}
\end{equation}
which defines $\gamma(t,\zeta, \bar \zeta)$. Equation \eqref{h2}
\begin{equation}
\partial_{\zeta} K+\beta=0\quad \text{and\quad c.c.}\label{h2III}
\end{equation}
defines  $\beta(t,\zeta)$, but also contrains $K$ since now 
\begin{equation}
\partial_{\bar\zeta} \partial_{\zeta} K=0.\label{h2IIIp}
\end{equation}
From the bulk perspective, the vanishing Bondi mass reads $\Psi_2=0$. Together with Eq. \eqref{h2IIIp}, these  are precisely the conditions for the Robinson--Trautman be Petrov type III (see \cite{Stephani:624239}). The principal null direction $\mathbf{k}$ is now triply degenerate. 

\item Set alternatively $M(t)=0$ and $\beta(t,\zeta)=0$. This amounts to keeping a pure-radiation reference energy--momentum tensor with $\gamma=\gamma(t,\zeta)$ due to the conservation equation \eqref{rmth}. The algebraic conditions \eqref{h3} and \eqref{h2} become
\begin{equation}
\partial_t\left(\frac{\partial^2_{\zeta} P}{P}\right) +\gamma=0\quad \text{and\quad c.c.} \label{h3II}
\end{equation}
and
\begin{equation}
\partial_{\zeta} K=0,\quad \partial_{\bar\zeta} K=0.\label{h2II}
\end{equation}
The first, Eq. \eqref{h3II}, defines $\gamma(t,\zeta)$. Since the latter does not depend on $\bar \zeta$, this imposes 
$\partial_{\bar\zeta}\partial_t\left(\frac{\partial^2_{\zeta} P}{P}\right)=0$ and its complex conjugate.  However, these are automatically satisfied by any $P(t,\zeta,\bar \zeta)$ satisfying  \eqref{h2II}.\footnote{The most general $P(t,\zeta,\bar \zeta)$ solving  \eqref{h2II} has been found in \cite{Foster:1967:NRT}. It reads:
\begin{equation}
\nonumber
P(t,\zeta,\bar \zeta)=\frac{1+\frac{\epsilon}{2} h(t,\zeta)\, \bar h(t,\bar\zeta)}{\sqrt{2f(t)\, \partial_\zeta h(t,\bar\zeta)\, \partial_{\bar\zeta}\bar h(t,\bar\zeta)}}
\end{equation}
with $\epsilon=0,\pm1$ and arbitrary functions $f(t)$ and $h(t,\zeta)$.}
Consequently, the only relevant conditions are the vanishing Bondi mass setting $\Psi_2=0$, and \eqref{h2II}
setting $\Psi_3=0$. With the present choice of boundary energy--momentum tensor the bulk is Robinson--Trautman Petrov type N and $\mathbf{k}$ is here quadruply degenerate. 

\item We can also choose a boundary reference energy--momentum tensor \eqref{gent} of pure perfect-fluid content, with shear-free congruences $\text{u}^{\pm}$. This latter condition is a free choice from our side,
which simply requires the coordinate dependence $\alpha^-(t,\zeta)= \left( \nicefrac{M(t) k^2}{8\pi G}\right)^{\nicefrac{1}{3}} \tilde \alpha^-(\zeta)$. The conditions \eqref{h3} and \eqref{h2} read:  
\begin{equation}
 \partial_t\left(\frac{\partial^2_{\zeta} P}{P}\right) +\frac{3}{2}Mk^4\frac{(\alpha^+)^2}{P^4}=0\quad \text{and\quad c.c. },\label{h3D}
\end{equation}
and 
\begin{equation}
\partial_{\zeta} K=3Mk^2\frac{\alpha^+}{P^2}\quad \text{and\quad c.c. }.\label{h2D}
\end{equation}
These define   $\alpha^-(t,\zeta)$ and  $\alpha^+(t,\bar \zeta)$ and set two conditions for $P$ and $K$:
\begin{equation}
6M \partial_t\left(\frac{\partial^2_\zeta P}{P}\right) + \left( \partial_\zeta K\right)^2=0\quad \text{and\quad c.c. },\label{D}
\end{equation}
plus 
\begin{equation}
 \partial_{\bar\zeta} \left(P^2 \partial_{\bar\zeta} K\right) 
 =0\quad \text{and\quad c.c. }.\label{D1}
\end{equation}
The latter stems out of the $\text{u}^{\pm}$-shearlessness condition, while the first is just the combination of  
\eqref{h3D} and \eqref{h2D}. From the bulk perspective, \eqref{D} and \eqref{D1} are nicely packaged in
\begin{equation}
3\Psi_2\Psi_4=2\Psi_3^2.
\end{equation}
With the null tetrad adopted for the bulk, this condition guarantees the Robinson--Trautman solution be Petrov type D. In this case, the principal null direction $\mathbf{k}$ remains doubly degenerate, while an extra doubly generate principal null direction emerges.

Two remarks are in order here. The first concerns the actual solutions of Robinson--Trautman Petrov type D. 
These are the Schwarzschild AdS and the $C$-metric AdS, which also belongs to the class of Pleba\'nski--Demia\'nski, with black-hole acceleration parameter. The  Pleba\'nski--Demia\'nski, without black-hole acceleration parameter has been obtained along the lines of though of the present work in \cite{Mukhopadhyay:2013gja}. The $C$-metric holography, has also been analyzed in \cite{Hubeny:2009kz,Caldarelli:2011wa}, and reveals many interesting peculiarities. The second remark is that a pure perfect-fluid reference energy--momentum tensor with arbitrary congruences $\text{u}^{\pm}$ would have led to the condition \eqref{D} only. Without \eqref{D1} the bulk would have been type II -- not the most general though. This apparent violation of the one-to-one correspondence between canonical classes of boundary tensors and  bulk Petrov types is due to the fact  that we are using the reference tensors $\text{T}^\pm$ instead of  $\text{S}^\pm$, as explained in Sec. \ref{Pet}. 

\item Finally, we can simply set $T^\pm=0$. This case is somehow degenerate. Indeed, according to \eqref{C-con}, the boundary has vanishing Cotton tensor and is thus conformally flat. So is the bulk since $\Psi_i=0$ for all $i=0,\ldots,4$. The bulk Robinson--Trautman is now Petrov type O, which reduces to pure four-dimensional anti-de Sitter spacetime. 
\end{itemize}

{
\subsection{Adding vorticity: towards Pleba\'nski--Demia\'nski} \label{vort}

The hydrodynamic congruence carries vorticity when allowing for non-trivial $\text{b}$ in \eqref{ut}. In this instance a genuine resummation operates in \eqref{papaefgenres} because $\rho \neq r$ (see \eqref{rho2}).
Boundary data of this kind were discussed in
 \cite{Mukhopadhyay:2013gja} together with their resummed exact ascendents, demonstrating the power of the  resummation. In the cases at hand,
the boundary metric is \eqref{PDbdymet} with $\Omega = 1$ and 
the vector $\partial_t$ tangent to the hydrodynamic congruence is assumed to be a Killing vector. This makes $\text{u}=-\text{d}t+\text{b}$ geodesic, shear- and expansion-free with vorticity 
$
\omega=\frac12\,\text{db}.
$
The reference tensors $\text{T}^\pm$ are chosen to be of the perfect-fluid form  $\text{T}^\pm_{\text{pf}}$ given in 
\eqref{RT-perflu} with equal velocity fields $\text{u}^+=\text{u}^-=\text{u}$. Being geodesic and expansion-free, 
they allow the conservation of this tensor with \emph{constant} $M_\pm$ (see App. \ref{appendix.perfect}): 
\begin{equation}
\label{perf-geo}
\text{T}^\pm_{\text{pf}}= \frac{k^2}{8\pi G}\left(m\pm i\frac{c}{2k^4}\right)\left(3\left(\text{d}t-\text{b}\right)^2 +\text{d}s^2\right).
\end{equation}
Imposing the conditions given by Eqs. \eqref{C-con} and \eqref{T-con}, this choice leads to (\romannumeral1) boundary 
geometries with perfect-fluid-like Cotton tensor, named \emph{perfect geometries} in \cite{Caldarelli:2012cm,Mukhopadhyay:2013gja} and 
(\romannumeral2) perfect-fluid physical boundary energy--momentum tensor. The latter statement shows that the boundary 
state is purely hydrodynamic with many vanishing transport coefficients, whereas the former leads to a family of boundary 
metrics depending on two real parameters, with hyperbolic, flat or spherical 
spatial parts.
%  $\nicefrac{2}{k^2P^2}\text{d}\zeta\text{d}\bar\zeta$. 

The resulting bulk geometry \eqref{papaefgenres} turns out to be the general AdS--Kerr--Taub--NUT black-hole spacetime with hyperbolic, 
flat or spherical  horizon, depending on three real parameters: the mass $m$,  the angular velocity $a$ and the nut charge 
$n$.\footnote{{Bulk angular velocity and nut charge act both as sources for boundary vorticity \cite{Leigh:2011au,Leigh:2012jv}.}}  
The nut charge and the angular velocity $a$ are encapsulated inside the constant $c$ in \eqref{perf-geo}.
These geometries belong to the most general Petrov D class of Einstein solutions
having two Killing vectors, namely the Pleba\'nski--Demia\'nski family \cite{Plebanski:1976gy}. 
As already mentioned when quoting the $C$-metric at the end of Sec. \ref{RRT}, the Pleba\'nski--Demia\'nski 
class has an extra physical parameter (the acceleration parameter), which can be introduced from the boundary 
perspective by relaxing the requirement $\text{u}^+=\text{u}^-$. The details of this case will appear elsewhere.

}

\section*{Conclusions}
\addcontentsline{toc}{section}{Conclusions}

In order to put our results in perspective, let us come back to the original question asked in the introduction:  \emph{given a class of boundary metrics, what are the conditions it should satisfy, and which energy--momentum tensor should it be accompanied with in order for an \emph{exact} dual bulk Einstein space to exist?}  Our answer to this question is based on three steps and four equations:
\begin{itemize}
\item The first step consists in choosing a set of two complex-conjugate reference tensors $\text{T}^\pm$, symmetric, traceless and satisfying the conservation equation \eqref{Tref-cons}.
\item Next, this tensor enables us  (\romannumeral1) to set conditions on the boundary metric by imposing its Cotton be the imaginary part of $\text{T}^\pm$ (up to constants), Eq. \eqref{C-con}; (\romannumeral2) to determine the boundary energy--momentum tensor as its real part, Eq. \eqref{T-con}.
\item Finally, using these data and Eq. \eqref{papaefgentetr}, we reconstruct the bulk Einstein space.
\end{itemize}

Several comments are in order here for making the picture complete. Equation \eqref{papaefgentetr} is obtained using the derivative expansion, which is an alternative to the Fefferman--Graham expansion and better suited for our purposes. As such, it assumes that the boundary state is in the hydrodynamic regime, described by an energy--momentum tensor of the fluid type. The latter has a natural built-in velocity field, interpreted as the fluid velocity congruence. Our method (first and second steps), however, does not necessarily lead to a fluid-like energy--momentum tensor. This is not a principle problem, because non-perturbative contributions with respect to the derivative expansion (non-hydrodynamic modes) are indeed expected to emerge along with a resummation \cite{Heller:2013fn}. In practice, though, it requires an extra piece of information regarding the velocity field around which the hydrodynamic modes are organized. 
To face this issue, we made the most economical choice, with a fluid at rest (Eq. \eqref{ut}) in the natural frames associated with the coordinates in use in the boundary metric \eqref{PDbdymet}. This choice is in agreement with the assumption of absence of shear, crucial for eliminating many terms in the derivative expansion of the bulk metric and making it resummable.

We have not formally proven that the three-step procedure proposed here leads indeed to Einstein spaces. 
However, our approach makes it clear that canonical boundary reference tensors guarantee the bulk be algebraically special. As a bonus, it is possible to set a precise relationship between the Segre type of the reference tensor and the Petrov type of the bulk Weyl tensor.
Many examples illustrate how the method works in practice and we have presented here the reconstruction of generic boundary data with a vorticity-free congruence. These lead to the whole family of Robinson--Trautman bulk Einstein spaces.

The formal proof of the constructive method presented in this paper will be released in the future. Besides that technical development, which we have chosen to avoid here, several other issues deserve further investigation. This effort aims at better understanding how the bulk is controlled from the boundary beyond any perturbative expansion. We know e.g. that the Petrov class of the bulk is determined by the choice of the boundary reference tensors. Our working assumption was the absence of shear for the boundary hydrodynamic congruence. 
Would shear be an obstruction to resummability? Can one reconstruct spaces which are not algebraically special, with zero shear on the boundary? Can one better understand the interplay between the two perturbative expansions mentioned here, namely the Fefferman--Graham and the derivative ones? Finally, based on the fact that Eqs. \eqref{Tref-cons} and \eqref{C-con} emerge as the boundary manifestation of Einstein's equations, we may wonder whether they possess some hidden symmetry \emph{\`a la} Geroch, which would relate integrable boundary data (see \cite{Leigh:2014dja} and the original references cited there).

\section*{Acknowledgements}

We would like to thank the University of Athens, the University of Bern, the \'Ecole Polytechnique and the University of Thessaloniki, where  the workshop \textsl{Aspects of fluid/gravity correspondence} was organized. All these institutions have provided hospitality and financial support during our collaboration meetings.
We also thank our colleagues M. Caldarelli, C. Charmousis, A. Gorantis, N. Halmagyi, R. Leigh and D. Klemm for useful discussions. The research of J.~Gath is supported by VILLUM FONDEN research grant VKR023371. 
Ayan Mukhopadhyay and A.C. Petkou are supported in part by European Union's Seventh Framework Programme under grant agreements (FP7-REGPOT-2012-2013-1) no 316165, the EU-Greece program \textsl{Thales} MIS 375734. Their research is also co-financed by the European Union (European Social Fund, ESF) and Greek national funds through the Operational Program \textsl{Education and Lifelong
Learning} of the National Strategic Reference Framework (NSRF) under \textsl{Funding
of proposals that have received a positive evaluation in the 3rd and 4th Call
of ERC Grant Schemes}. The work of A.C. Petkou is partially supported by the research grant ARISTEIA II, 3337, 
\textsl{Aspects of three-dimensional CFTs}, by the Greek General Secretariat of Research and Technology, and also  by the CreteHEPCosmo-228644 grant. The research of K.~Siampos is supported by the Swiss National Science Foundation.
Jakob Gath, P.M. Petropoulos and K.~Siampos acknowledge the \textsl{Germaine de Stael}
francoswiss bilateral program 2015 (project no 32753SG) for financial support.

\appendix

\section{On perfect-fluid dynamics}
%\addcontentsline{toc}{section}{On perfect-fluid dynamics}
\label{appendix.perfect}

In this appendix, we would like to set a useful criterion regarding the motion of conformal perfect fluids. For such fluids with velocity congruence $\text{u}$ and pressure $p(x)$, three-dimensional Euler's equations read:
\begin{equation}
\label{PMP-Euler0}
 \begin{cases}
2\text{u}(p)+3p\, \Theta=0, \\
 \text{u}(p)\, \text{u}
+\text{d}p+3p\, \mathrm{a}=0.
\end{cases}
\end{equation}
where $\text{u}(p)=u^\mu \partial_\mu p$.
Combining these equations, we obtain:
\begin{equation}\label{Euler0-int}
3\text{A}+\frac{\text{d}p}{p}=0,
\end{equation}
where $\text{A}=\text{a} -\frac{\Theta}{2} \text{u}$.
For Eq. \eqref{Euler0-int} to hold we extract a simple integrability condition: the Weyl connection $\text{A}$ must be closed (hence locally exact) for a pressure field $p(x)$ to exist and account for the expansion and acceleration of the fluid. If  $\text{A}$ vanishes, the pressure is constant; if 
$\text{A}$ is not exact, the fluid moving on the congruence $\text{u}$ is not perfect, or even the hydrodynamic regime is not applicable.

%\addcontentsline{toc}{section}{References}
%\bibliographystyle{newutphys}
%\bibliography{References}

%\input{referenc}
\end{document}